\documentclass{sig-alternate}

\usepackage{framed}
\usepackage{booktabs}
\usepackage{subfigure}
\usepackage{comment}
\usepackage{mdwlist}

\newcommand{\mysec}[1]{Section~\ref{sec:#1}}

\newcommand{\myeq}[1]{Equation~\ref{eq:#1}}

\newcommand{\myfig}[1]{Figure~\ref{fig:#1}}

\newcommand{\g}{\,\vert\,}
\newcommand{\E}{\textrm{E}}

\newcommand{\poisson}{\textrm{Poisson}}
\newcommand{\mult}{\textrm{Mult}}

\newcommand{\gam}{\textrm{Gamma}}
\newcommand{\shape}{\textrm{shp}}
\newcommand{\rate}{\textrm{rte}}

\begin{document}
%

\title{Scalable Recommendation with Poisson Factorization}

%
%
%
%
%

\numberofauthors{3}

%
\author{
%
%
\alignauthor
Prem Gopalan\\
\affaddr{Princeton University}\\
\affaddr{35 Olden Street}\\
\affaddr{Princeton, NJ}\\
\email{pgopalan@cs.princeton.edu}
\alignauthor
Jake M. Hofman \\
\affaddr{Microsoft Research}\\
\affaddr{641 Sixth Avenue, Floor 7}\\
\affaddr{New York, NY}\\
\email{jmh@microsoft.com}
\alignauthor
David M. Blei \\
\affaddr{ Princeton University}\\
\affaddr{35 Olden Street}\\
\affaddr{Princeton, NJ}\\
\email{blei@cs.princeton.edu}
}
\maketitle


\begin{abstract}
We develop hierarchical Poisson matrix factorization (HPF) for
recommendation.  HPF models sparse user behavior data, large user/item
matrices where each user has provided feedback on only a small subset
of items.  HPF handles both explicit ratings, such as a number of
stars, or implicit ratings, such as views, clicks, or purchases.  We
develop a variational algorithm for approximate posterior inference
that scales up to massive data sets, and we demonstrate its
performance on a wide variety of real-world recommendation
problems--users rating movies, users listening to songs, users reading
scientific papers, and users reading news articles.  Our study reveals
that hierarchical Poisson factorization definitively outperforms
previous methods, including nonnegative matrix factorization, topic
models, and probabilistic matrix factorization
techniques.
\end{abstract}




\section{Introduction}

Recommendation systems are a vital component of the modern Web.  They
help readers effectively navigate otherwise unwieldy archives of
information and help websites direct users to items---movies,
articles, songs, products---that they will like.
A recommendation system is built from user behavior data, historical
data about which items each user has consumed, be it clicked, viewed,
rated, or purchased. First, we uncover the behavioral patterns that
characterize various types of users and the kinds of items they tend
to like.  Then, we exploit these discovered patterns to recommend
future items to its users.

In this paper, we develop Poisson factorization (PF) algorithms for
recommendation.  Our algorithms easily scale to massive data and
significantly outperform the existing methods.  We show that
Poisson factorization for recommendation is tailored to real-world
properties of user behavior data: the heterogenous interests of users,
the varied types of items, and a realistic distribution of the finite
resources that users have to consume items.

\begin{figure*}[th]
\centering
\vspace{0.1cm}
\small
\begin{tabular}{c}
\toprule
\bf{``Action''}\\
\midrule
The Matrix\\
The Matrix: Reloaded\\
Spider-Man\\
X2: X-Men United\\
\bottomrule
\end{tabular}
\begin{tabular}{c}
\toprule
\bf{``Indie Comedy, Romance''}\\
\midrule
Grosse Pointe Blank\\
Four Weddings and a Funeral\\
High Fidelity\\
Much Ado About Nothing\\
\bottomrule
\end{tabular}
\begin{tabular}{c}
\toprule
\bf{``80's Science Fiction''}\\
\midrule
Star Wars: Episode IV: A New Hope\\
Star Wars: Episode VI: Return of the Jedi\\
Star Wars: Episode V: The Empire Strikes Back\\
Back to the Future Part II\\
\bottomrule
\end{tabular}

\vspace{0.5cm}

\centering
\includegraphics[width=\textwidth]{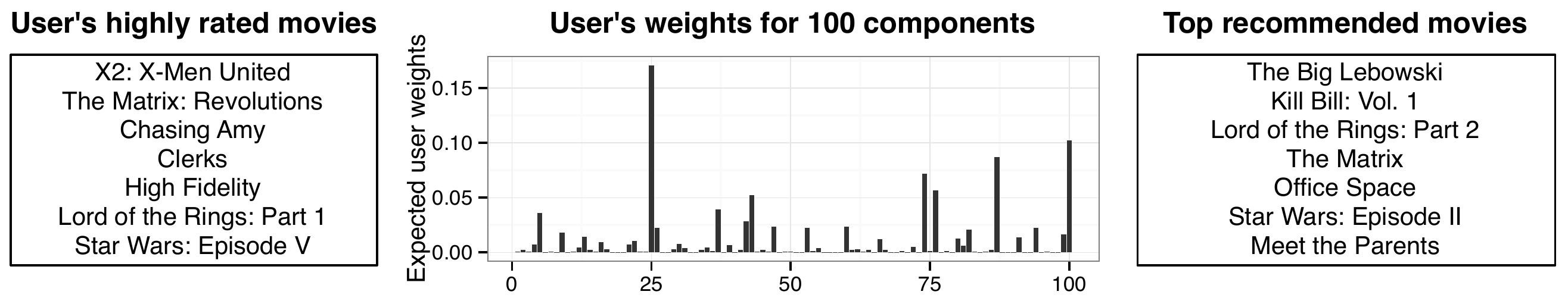}\\
\caption{ The top panel shows the top movies in 3 components for a
  user from the Netflix data set. The bottom panel is an illustration
  showing a subset of the highly rated movies by this user, and the
  right panel shows movies recommended to the user by our
  algorithm. The expected user's $K$-vector of weights $\theta_u$,
  inferred by our algorithm is shown in the middle panel.}
\label{fig:netflix-illustration}
\end{figure*}

\myfig{netflix-illustration} illustrates Poisson factorization on data
from Netflix.  The Netflix data contains the ratings of 480,000 users
on 17,000 movies, organized in a matrix of 8.16B cells (and containing
250M ratings).  From these data, we extract the patterns of users'
interests and the movies that are associated with those interests.
The left panel illustrates some of those patterns---the algorithm has
uncovered action movies, independent comedies, and 1980s science
fiction.

The top panel illustrates how we can use these patterns to form
recommendations for an (imaginary) user.  This user enjoys various
types of movies, including fantasy (``Lord of the Rings''), classic
science fiction (``Star Wars: Episode V''), and independent comedies
(``Clerks'', ``High Fidelity'').  Of course, she has only seen a
handful of the available movies.  PF first uses the movies she has
seen to infer what kinds of movies she is interested in, and then uses
these inferred interests to suggest new movies.  The
list of movies at the bottom of the figure was suggested by our
algorithm. It includes other comedies (such as ``The Big Lebowski'') and
other science fiction (such as ``Star Wars: Episode II'').

In more detail, Poisson factorization is a probabilistic model of
users and items.  It associates each user with a latent vector of
preferences, each item with a latent vector of attributes, and
constrains both sets of vectors to be sparse and non-negative.  Each
cell of the observed behavior matrix is assumed drawn from a Poisson
distribution---an exponential family distribution over non-negative
integers---whose parameter is a linear combination of the
corresponding user preferences and item attributes.  The main
computational problem is posterior inference: given an observed matrix
of user behavior, we discover the latent attributes that describe the
items and the latent preferences of the users.  For example, the
components in \myfig{netflix-illustration} (left) illustrate the top
items for specific attribute dimensions and the plot in
\myfig{netflix-illustration} (middle) illustrates the estimated
preference vector for the given user.  A spike in the preference
vector implies that the user tends to like items with the
corresponding latent attribute.

This general procedure is common to many variants of matrix factorization.
We found,
however, that PF enjoys significant quantitative advantages over
classical methods and for a wide variety of data sets, including those
with implicit feedback (a binary matrix indicating which items users
consumed) and those with explicit feedback (a matrix of integer
ratings).  \myfig{precision_recall} shows that PF, and its
hierarchical variant HPF, perform significantly better than
existing methods---including the industry standard of matrix
factorization with user and item biases (MF)---for large data sets of
Netflix users watching movies, Last.FM users listening to music,
scientists reading papers, and \textit{New York Times} readers
clicking on articles.


There are two main advantages of Poisson factorization over
traditional methods, both of which contribute to its superior
empirical performance.  First, it better captures real consumption
data, specifically that users have finite (and varied) resources with
which to view items.  To see this, we can rewrite the model as a two
stage process where a user first decides on a budget of movies to
watch and then spends this budget watching movies that she is
interested in.  If the model accurately captures the distribution of
budgets then watched items carry more weight than unwatched items,
because unwatched items can be partially explained by a lack of
resources. We conjecture that classical matrix factorization
systematically overestimates the users' budgets, and we confirm this
hypothesis in \mysec{eval} using a posterior predictive
check~\cite{Gelman:1996}.  This misfit leads to an overweighting of
the zeros, which explains why practitioners require complex methods
for downweighting
them~\cite{Hu:2008p9402,Gantner:2012p9364,Dror:2012a,Paquet:2013p9197}.
Poisson factorization does not need to be modified in this way.

The second advantage of PF algorithms is that they need only iterate
over the viewed items in the observed matrix of user behavior, i.e.,
the non-zero elements, and this is true even for implicit or
``positive only'' data sets.  (This follows from the mathematical form
of the Poisson distribution.)  Thus, Poisson factorization takes
advantage of the natural sparsity of user behavior data and can easily
analyze massive real-world data. In contrast, classical matrix
factorization based on the Gaussian
distribution~\cite{Salakhutdinov:2008} must iterate over both positive
and negative examples in the implicit setting. Thus it cannot take
advantage of data sparsity, which makes computation difficult for even
modestly sized problems.  For example, one cannot fit to the full
Netflix data set (as we did in \myfig{netflix-illustration}) without appealing to
stochastic optimization~\cite{Mairal:2010}.  We note that our
algorithms are also amenable to stochastic optimization, which we can
use to analyze data sets even larger than those we studied.


We review related work below before discussing details of the Poisson
factorization model, including its statistical properties and methods
for scalable inference.

\section{Related work}
The roots of Poisson factorization come from nonnegative matrix
factorization~\cite{Lee:1999}, where the objective function is
equivalent to a factorized Poisson likelihood.  The original NMF
update equations have been shown to be an expectation-maximization
(EM) algorithm for maximum likelihood estimation of a Poisson model
via data augmentation~\cite{Cemgil:2009}.

Placing a Gamma prior on the user weights results in the GaP
model~\cite{Canny:2004}, which was developed as an alternative text
model to latent Dirichlet allocation
(LDA)~\cite{Blei:2003b,Inouye:2014}. The GaP model is fit using the
expectation-maximization algorithm to obtain point estimates for user
preferences and item attributes. The Probabilistic Factor Model
(PFM)~\cite{Ma:2011} improves upon GaP by placing a Gamma prior on the
item weights as well, and using multiplicative update rules to infer
an approximate maximum a posteriori estimate of the latent factors.
In contrast, as explained below, our model uses a hierarchical prior
structure of Gamma priors on user and item weights, and Gamma priors
over the rate parameters from which these weights are drawn. This
enables us to accurately model the skew in user activity and item
popularity, which contributes to good predictive
performance. Furthermore, we approximate the full posterior over all
latent factors using a scalable variational inference algorithm.

Independently of GaP and user behavior models, Poisson factorization
has been studied in the context of signal processing for source
separation~\cite{Cemgil:2009,Hoffman:2012} and for the purpose of
detecting community structure in network
data~\cite{Ball:2011,Gopalan:2013}. This research includes variational
approximations to the posterior, though the issues and details around
these data differ significantly from user data we consider and our
derivation below (based on auxiliary variables) is more direct.

When modeling implicit feedback data sets, researchers have proposed
merging factorization techniques with neighborhood
models~\cite{Koren:2008}, weighting techniques to adjust the relative
importance of positive examples~\cite{Hu:2008p9402}, and
sampling-based approaches to create informative negative
examples~\cite{Gantner:2012p9364,Dror:2012a,Paquet:2013p9197}.  In
addition to the difficulty in appropriately weighting or sampling
negative examples, there is a known selection bias in provided ratings
that causes further complications~\cite{Marlin:2009,Marlin:2012}.
Poisson factorization does not require such special adjustments and
scales linearly with the number of observed ratings.

We discuss additional related recommendation methods in
\mysec{eval}, where we compare a variety of applicable methods to
Poisson factorization empirically.

\section{Poisson Recommendation}
\label{sec:model}


In this section we describe the Poisson factorization model for
recommendation, and discuss its statistical properties.

We are given data about users and items, where each user has consumed
and possibly rated a set of items.  The observation $y_{ui}$ is the
rating that user $u$ gave to item $i$, or zero if no rating was given.
(In so-called ``implicit'' consumer data, $y_{ui}$ equals one if user
$u$ consumed item $i$ and zero otherwise.)  User behavior data, such
as purchases, ratings, clicks, or views, are typically sparse.  Most
of the values of the matrix $y$ are zero.

We model these data with factorized Poisson
distributions~\cite{Canny:2004}, where each item $i$ is represented by
a vector of $K$ latent attributes $\beta_i$ and each user $u$ by a
vector of $K$ latent preferences $\beta_u$.  The observations $y_{ui}$
are modeled with a Poisson, parameterized by the inner product of the
user preferences and item attributes, $y_{ui} \sim
\poisson(\theta_u^\top \beta_i)$.  This is a variant of probabilistic
matrix factorization~\cite{Salakhutdinov:2008a} but where each user
and item's weights are positive~\cite{Lee:1999} and where the Poisson
replaces the Gaussian.

Beyond the basic data generating distribution, we place Gamma priors
on the latent attributes and latent preferences, which encourage the
model towards sparse representations of the users and items.
Furthermore, we place additional priors on the user and item-specific
rate parameter of those Gammas, which controls the average size of the
representation.  This hierarchical structure allows us to capture the
diversity of users, some tending to consume more than others, and the
diversity of items, some being more popular than others.  The
literature on recommendation systems suggests that a good model must
capture such heterogeneity across users and items~\cite{Koren:2009}.

Putting this together, the generative process of the hierarchical
Poisson factorization model (HPF) is as follows:
\begin{enumerate*}
\item For each user $u$:
  \begin{enumerate*}
  \item Sample activity $\xi_u \sim \gam(a', a'/b')$.
  \item For each component $k$, sample preference $$\theta_{uk} \sim
    \gam(a, \xi_u).$$
  \end{enumerate*}

\item For each item $i$:
  \begin{enumerate*}
    \item Sample popularity $\eta_i \sim \gam(c', c'/d')$.
    \item For each component $k$, sample attribute
      $$\beta_{ik} \sim \gam(c, \eta_i).$$
  \end{enumerate*}

\item For each user $u$ and item $i$, sample rating
  $$y_{ui} \sim
  \poisson(\theta_u^\top \beta_i).$$
\end{enumerate*}
This process describes the statistical assumptions behind the model.
Note that we also study a sub-class of HPF where we fix the rate
parameters for all users and items to the same pair of
hyperparameters. We call this model Bayesian Poisson Factorization
(BPF).

The central computational problem is posterior inference, which is
akin to ``reversing'' the generative process.  Given a user behavior
matrix, we want to estimate the conditional distribution of the latent
per-user and per-item structure, $p(\theta_{1:N} \beta_{1:M} \g y)$.
The posterior is the key to recommendation.  We estimate the posterior
expectation of each user's preferences, each items attributes and,
subsequently, form predictions about which unconsumed items each user
will like.  We discuss posterior inference in
\mysec{inference}.


Once the posterior is fit, we use HPF to recommend items to users by
predicting which of the unconsumed items each will like.  We rank each
user's unconsumed items by their posterior expected Poisson
parameters,
\begin{equation}
  \label{eq:score}
  \textrm{score}_{ui} = \E[\theta_u^\top \beta_i \g y].
\end{equation}
This amounts to asking the model to rank by probability which of the
presently unconsumed items each user will likely consume in the
future.

\subsection{Properties of HPF}
With the modeling details in place, we highlight several statistical
properties of hierarchical Poisson factorization.  These properties
provide advantages over classical (Gaussian) matrix
factorization.\footnote{Specifically, by classical matrix
  factorization we mean L2 regularized matrix factorization with
  bias terms for users and items, fit using stochastic gradient
  descent~\cite{Koren:2009}. Without the bias terms, this corresponds
  to maximum a-posteriori inference under Probabilistic Matrix
  Factorization~\cite{Salakhutdinov:2008a}.}

{\bf HPF captures sparse factors}.  As we mentioned above, the Gamma
priors on preferences and attributes encourages sparse representations
of users and items.  Specifically, by setting the shape parameter to
be small, most of the weights will be close to zero and only a few
will be large.


{\bf HPF models the long-tail of users and items}.  One statistical
characteristic of real-world user behavior data is the distribution of
user activity (i.e., how many items a user consumed) and item
popularity (i.e., how many users consumed an item).  These
distributions tend to be long-tailed: while most users consume a
handful few items, a few ``tail users'' consume thousands of items.  A
question we can ask of a statistical model of user behavior data is
how well it captures these distributions.  We found that HPF captures
them very well, while classical matrix factorization does not.

To check this, we implemented a \textit{posterior predictive check}
(PPC)~\cite{Rubin:1984,Gelman:1996}, a technique for model assessment
from the Bayesian statistics literature.  The idea behind a PPC is to
simulate a complete data set from the posterior predictive
distribution---the distribution over data that the posterior
induces---and then compare the generated data set to the true
observations. A good model will produce data that captures the
important characteristics of the observed data.


We developed a PPC for matrix factorization algorithms on user
behavior data.  First, we formed posterior estimates of user
preferences and item attributes for both classical MF and HPF.  Then,
from these estimates, we simulated user behavior by
drawing values for each user and item.  (For classical matrix
factorization, we truncated these values at zero and rounded to one in
order to generate a plausible matrix.)  Finally, we compared
the matrix generated by the posterior predictive distribution to the
true observations.

\begin{figure}[t!]
  \centering
  \includegraphics[width=0.4\textwidth]{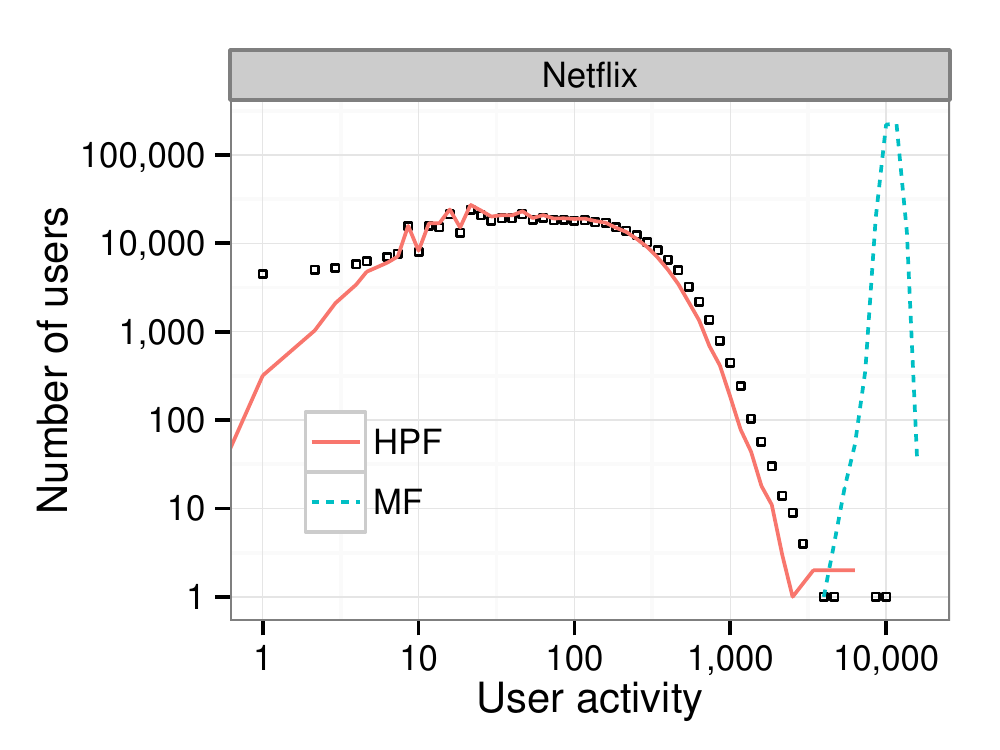}
  \caption{A posterior predictive check of the distribution of total
    ratings for the Netflix data set.  The pink curve shows the
    empirical count of the number of users who have rated a given
    number of items, while the green and blue curves show the
    simulated totals from fitted Poisson and traditional matrix
    factorization models, respectively. The Poisson marginal closely
    matches the empirical, whereas classical matrix factorization fits
    a large mean to account for skew in the distribution and the
    missing ratings.}
\label{fig:marginals}
\end{figure}

\myfig{marginals} illustrates our PPC for the Netflix data.  In this
figure, we illustrate three distributions over user activity: the
observed distribution (squares), the distribution from a data set
replicated by HPF (red line), and a distribution from a data set
replicated by classical MF (blue line).  HPF captures the truth much
more closely than classical MF, which badly overestimates the
distribution of user activity.  (We note that this is true for the
item popularities as well, and for the other data sets.) This
indicates that HPF better represents real data when measured by its
ability to capture distributions of user activity and item popularity.



{\bf HPF downweights the effect of zeros.}  Another advantage of HPF
is that it implicitly down-weights the contribution of the items that
each user did not consume.  With an appropriate fit to user activity,
the model has two ways of explaining an unconsumed item: either the
user is not interested in it or she would be interested in it but is
likely to not be further active. In contrast, a user that consumes an
item must be interested in it.  Thus, the model benefits more from
making a consumed user/item pair more similar than making an
unconsumed user/item pair less similar.

Classical MF is based on Gaussian likelihoods (i.e., squared loss),
which gives equal weight to consumed and unconsumed items.
Consequently, when faced with a sparse matrix and implicit feedback,
i.e., binary consumption data, matrix factorization places more total
emphasis on the unconsumed user/item pairs.  (This too can be seen to
stem from classical MF's overestimation of the distribution of user
activity.)  To address this, researchers have patched MF in
complex ways, for example, by including per-observation
confidences~\cite{Koren:2009} or considering all zeroes to be hidden
variables~\cite{Paquet:2013p9197}.  Poisson factorization more
naturally solves this problem by better capturing user activity.

As an example, consider two similar science fiction movies, ``Star
Wars'' and ``The Empire Strikes Back'', and consider a user who has
seen one of them.  The Gaussian model pays an equal penalty for making
the user similar to these items as it does for making the user
different from them---with quadratic loss, seeing ``Star Wars'' is
evidence for liking science fiction, but not seeing ``The Empire
Strikes Back'' is evidence for disliking it.  The Poisson model,
however, will prefer to bring the user's latent weights closer to the
movies' weights because it favors the information from the user
watching ``Star Wars''. Further, because the movies are similar, this
increases the Poisson model's predictive score that a user who watches
``Star Wars'' will also watch ``The Empire Strikes Back''.

{\bf Fast inference with sparse matrices.}  Finally, the likelihood of
the observed data under HPF (and BPF) depends only on the consumed
items, that is, the non-zero elements of the user/item matrix $y$.
This facilates computation for the kind of sparse matrices we
observe in real-world data.

We can see this property from the form of the Poisson distribution.
Given the latent preferences $\theta_u$ and latent attributes
$\beta_i$, the Poisson distribution of the rating $y_{ui}$ is
\begin{equation}
  p(y_{ui} \g \theta_u, \beta_i) =
  \left(\theta_u^\top \beta_i\right)^y
  \exp\left\{-\theta_u^\top \beta_i \right\} / y_{ui}!
\end{equation}
Recall the elementary fact that $0! = 1$.  The log probability of the
complete matrix $y$ is
\begin{align}
  \log p(y \g \theta, \beta) =
  & \left(\textstyle \sum_{\{y_{ui} > 0\}}
    y_{ui} \log (\theta_u^\top \beta_i) - \log y_{ui}!
  \right) \\
  & -
  \left(\textstyle\sum_{u} \theta_u\right)^\top \left(\textstyle
    \sum_{i} \beta_i\right). \nonumber
\end{align}

Classical MF does not enjoy this property. These methods, especially
when applied to massive data sets of implicit feedback, must (in
theory) iterate over all the cells of the matrix.  Practitioners
require solutions such as sub-sampling~\cite{Dror:2012a}
approximation~\cite{Hu:2008p9402}, or stochastic
optimization~\cite{Mairal:2010}.

\subsection{Inference with variational methods}
\label{sec:inference}

Using HPF for recommendation hinges on solving the posterior inference
problem.  Given a set of observed ratings, we would like to infer the
user preferences and item attributes that explain these ratings, and
then use these inferences to recommend new content to the users.  In
this section we discuss the details and practical challenges of
posterior inference for HPF, and present a mean-field variational
inference algorithm as a practical and scalable approach.  Our
algorithm easily accommodates data sets with millions of users and
hundreds of thousands of items on a single CPU.

Given a matrix of user behavior, we would like to
compute the posterior distribution of user preferences $\theta_{uk}$,
item attributes $\beta_{ik}$, user activity $\xi_{u}$ and item
popularity $\eta_i$.  As for many Bayesian models, however, the exact
posterior is computationally intractable. We show how to
efficiently approximate the posterior with mean-field variational
inference.


Variational inference is an optimization-based strategy for
approximating posterior distributions in complex probabilistic
models~\cite{Jordan:1999,Wainwright:2008}.  Variational
algorithms posit a family of distributions over the hidden variables,
indexed by free ``variational'' parameters, and then find the member
of that family that is closest in Kullback-Liebler (KL) divergence to
the true posterior.  (The form of the family is chosen to make this
optimization possible.)  Thus, variational inference turns the
inference problem into an optimization problem.  Variational inference
tends to scale better than alternative sampling-based approaches, like
Monte Carlo Markov chain sampling, and has been deployed to solve many
applied problems with complex models, including large-scale
recommendation~\cite{Paquet:2013p9197}.


We will describe a simple variational inference algorithm for HPF.  To
do so, however, we first give an alternative formulation of the model
in which we add an additional layer of latent variables.  These
auxiliary variables facilitate derivation and description of the
algorithm~\cite{Ghahramani:2001,Hoffman:2013}.

For each user and item we add $K$ latent variables $z_{uik} \sim
\poisson(\theta_{uk} \beta_{ik})$, which are integers that sum to the
user/item value $y_{ui}$.  A sum of Poisson random variables is itself
a Poisson with rate equal to the sum of the rates.  Thus, these new
latent variables preserve the marginal distribution of the
observation, $y_{ui} \sim \poisson(\theta_{u}^\top \beta_{i})$.  These
variables can be thought of as the contribution from component $k$ to
the total observation $y_{ui}$.  Note that when $y_{ui} = 0$, these
auxiliary variables are not random---the posterior distribution of
$z_{ui}$ will place all its mass on the zero vector.  Consequently,
our inference procedure need only consider $z_{ui}$ for those
user/item pairs where $y_{ui} > 0$.

\begin{figure}[th]
  \begin{framed}
    For all users and items, initialize the user parameters
    $\gamma_u$, $\kappa_u^{\rate}$ and item parameters $\lambda_i$,
    $\tau_i^{\rate}$ to the prior with a small random offset. Set the
    user activity and item popularity shape parameters:
    \begin{align}
      \kappa_u^{\shape} = a' + Ka; \quad \tau_i^{\shape} = c' + Kc\nonumber
    \end{align}


    Repeat until convergence:
    \begin{enumerate}
    \item For each user/item such that $y_{ui} > 0$, update the multinomial:
      \begin{equation*}
        \phi_{ui} \propto \exp\{\Psi(\gamma_{uk}^\shape) - \log
        \gamma_{uk}^{\rate} + \Psi(\lambda_{ik}^\shape) - \log
        \lambda_{ik}^\rate\}.
      \end{equation*}
    \item For each user, update the user weight and activity parameters:
      \begin{align}
        \gamma_{uk}^\shape & = a + \textstyle \sum_{i} y_{ui}
        \phi_{uik} \nonumber\\
        \gamma_{uk}^\rate & = \frac{\kappa_u^{\shape}}{\kappa_u^{\rate}} + \textstyle \sum_i \lambda_{ik}^{\shape} / \lambda_{ik}^{\rate}\nonumber\\
        \kappa_{u}^\rate & = \frac{a'}{b'} + \sum_k \frac{\gamma_{uk}^{\shape}}{\gamma_{uk}^{\rate}}\nonumber
      \end{align}
    \item For each item, update the item weight and popularity parameters:
      \begin{align}
        \lambda_{ik}^\shape & = c + \textstyle \sum_{u} y_{ui}
        \phi_{uik}\nonumber\\
        \lambda_{ik}^\rate & = \frac{\tau_i^{\shape}}{\tau_i^{\rate}} + \textstyle \sum_u
        \gamma_{uk}^{\shape} / \gamma_{uk}^{\rate}\nonumber\\
        \tau_{i}^\rate & = \frac{c'}{d'} + \sum_k \frac{\lambda_{ik}^{\shape}}{\lambda_{ik}^{\rate}}\nonumber
      \end{align}
    \end{enumerate}
    \vspace{-0.2in}
\end{framed}
    \vspace{-0.2in}
\caption{\label{fig:batch}Variational inference for Poisson
  factorization.  Each iteration only needs to consider the non-zero
  elements of the user/item matrix.}
\end{figure}

With these latent variables in place, we now describe the algorithm.
First, we posit the variational family over the hidden variables.
Then we show how to optimize its parameters to find the member close
to the posterior of interest.

The latent variables in the model are user weights $\theta_{uk}$, item
weights $\beta_{ik}$, and user-item contributions $z_{uik}$, which we
represent as a $K$-vector of counts $z_{ui}$.  The \textit{mean-field
  family} considers these variables to be independent and each
governed by its own distribution,
\begin{eqnarray*}
  \label{eq:q}
  q(\beta, \theta, \xi, \eta, z) =& \prod_{i,k} q(\beta_{ik} \g \lambda_{ik})
  \prod_{u,k} q(\theta_{uk} \g \gamma_{uk}) \nonumber\\
  & \prod_{u} q(\xi_u \g \kappa_u) \prod_{i} q(\eta_i \g \tau_i)
  \prod_{u,i} q(z_{ui} \g \phi_{ui}).
\end{eqnarray*}
Though the variables are independent, this is a flexible family of
distributions because each variable is governed by its own free
parameter.  The variational factors for preferences $\theta_{uk}$,
attributes $\beta_{ik}$, activity $\xi_u$, and popularity $\eta_i$ are
all Gamma distributions, with freely set scale and rate variational
parameters. The variational factor
for $z_{ui}$ is a free multinomial, i.e., $\phi_{ui}$ is a $K$-vector
that sums to one.  This form stems from $z_{ui}$ being a bank of
Poisson variables conditional on a fixed sum $y_{ui}$, and the
property that such conditional Poissons are distributed as a
multinomial~\cite{Johnson:2005, Cemgil:2009}.


After specifying the family, we fit the variational parameters $\nu =
\{\lambda, \gamma, \kappa, \tau, \phi\}$ to minimize the KL divergence
to the posterior, and then use the corresponding variational
distribution $q(\cdot \g \nu^*)$ as its proxy. The mean-field
factorization facilitates both optimizing the variational objective
and downstream computations with the approximate posterior, such as
the recommendation score of \myeq{score}.


We optimize the variational parameters with a coordinate ascent
algorithm, iteratively optimizing each parameter while holding the
others fixed.  The algorithm is illustrated in \myfig{batch}. We
denote shape with the superscript ``shp'' and rate with the
superscript ``rte''.  (We omit a detailed derivation due to space constraints.)

Note that our algorithm is very efficient on sparse matrices. In step
1, we need only update variational multinomials for the non-zero
user/item observations $y_{ui}$.  In steps 2 and 3, the sums over
users and items need only to consider non-zero observations.  This
efficiency is thanks the likelihood of the full matrix only depending
on the non-zero observations, as we discussed in the previous section.




We terminate the algorithm when the variational distribution
converges. Convergence is measured by computing the prediction
accuracy on a validation set.  Specifically, we approximate the
probability that a user consumed an item using the variational
approximations to posterior expectations of $\theta_u$ and $\beta_i$,
and compute the average predictive log likelihood of the validation
ratings. The HPF algorithm stops when the change in log likelihood is
less than 0.0001\%. For the HPF and the BPF we find that the algorithm
is largely insensitive to small changes in the hyper-parameters. To
enforce sparsity, we set the shape hyperparameters $a'$, $a$, $c$ and
$c'$ to provide exponentially shaped prior \gam~distributions. We
fixed each hyperparameter at $0.3$. We set the hyperparameters $b'$
and $d'$ to 1, fixing the prior mean at 1.



\begin{figure*}[t!]
\centering
\includegraphics[width=\textwidth]{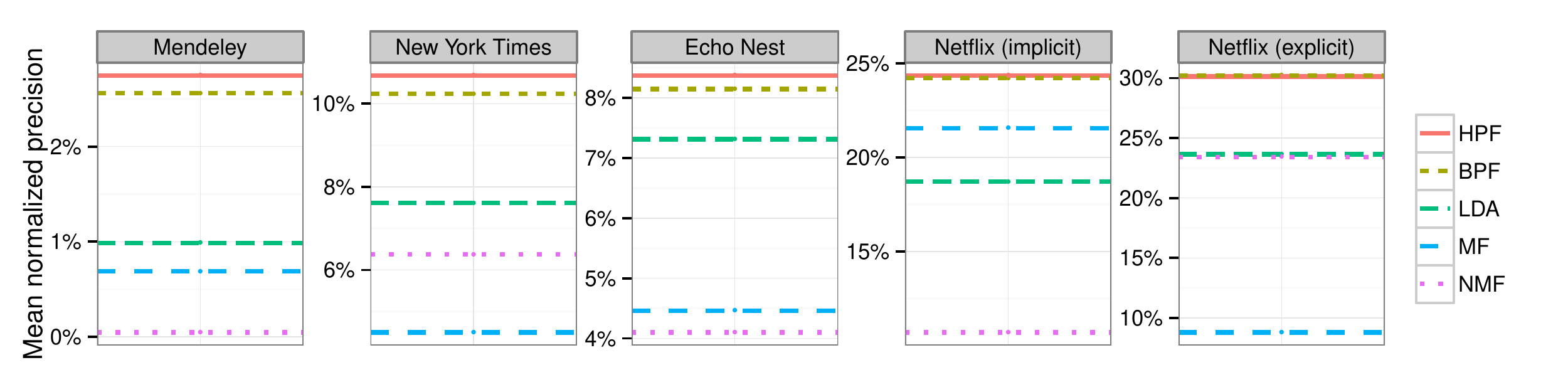}\\
\includegraphics[width=\textwidth]{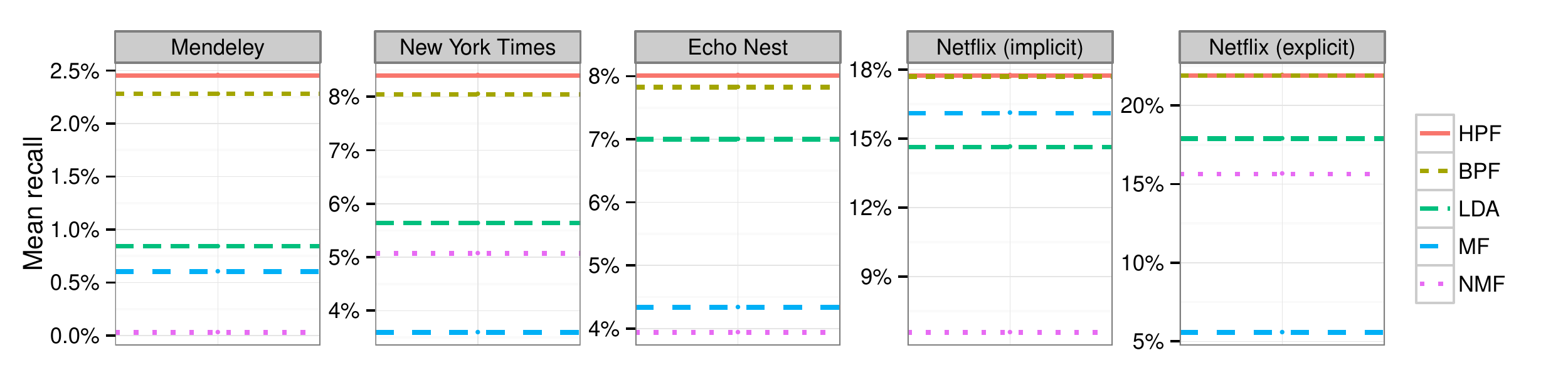}\\
\caption{Predictive performance on data sets. The top and bottom plots
  show normalized mean precision and mean recall at 20
  recommendations, respectively. While competing method performance varies
  across data sets, HPF and BPF consistently outperform competing
  methods.}
\label{fig:precision_recall}
\end{figure*}

\section{Empirical Study}
\label{sec:eval}
We evaluate the performance of the Hierarchical Poisson factorization
(HPF) algorithm and its non-hierarchical variant (BPF) on a variety of
large-scale user behavior data sets: users listening to music, users
watching movies, users reading scientific articles, and users reading
the newspaper.  We find that HPF and BPF give significantly better
recommendations than competing methods.

We first discuss the details of each data set and of the competing
recommendation methods. We then describe our study, noting the
superior performance and computational efficiency of HPF.  We conclude
with an exploratory analysis of preferences and attributes on several
of the data sets.

{\bf Data Sets.} We study the HPF algorithm in Figure~\ref{fig:batch}
on several data sets of user behavior, including both implicit and
explicit feedback:
\begin{itemize}
\item The {\bf Mendeley} data set~\cite{Jack:2010} of scientific
  articles is a binary matrix of 80,000 users and 260,000 articles,
  with 5 million observations.  Each cell corresponds to the presence
  or absence of an article in a scientist's online library.
\item The {\bf Echo Nest} music data set~\cite{Bertin-Mahieux:2011} is
  a matrix of 1 million users and 385,000 songs, with 48 million
  observations.  Each observation is the number of times a user played
  a song.
\item The {\bf New York Times} data set is a matrix of 1,615,675 users
  and 103,390 articles, with 80 million observations.  Each
  observation is the number of times a user viewed an article.

\item The {\bf Netflix} data set~\cite{Koren:2009} contains 480,000
  users and 17,770 movies, with 100 million observations. Each
  observation is the rating (from 1 to 5 stars) that a user provided
  for a movie.
\end{itemize}

The scale and diversity of these data sets enables a robust evaluation
of our algorithm. The Mendeley, Echo Nest, and New York Times data
are sparse compared to Netflix. For example, we observe
only 0.001\% of all possible user-item ratings in Mendeley, while 1\%
of the ratings are non-zero in the Netflix data. This is partially a
reflection of large number of items relative to number users in these
data sets.

Furthermore, the intent signaled by an observed rating varies
significantly across these data sets. For instance, the Netflix data
set gives the most direct measure of stated preferences for items, as
users provide an explicit star rating for movies they have watched. In
contrast, article click counts in the New York Times data are a less
clear measure of how much a user likes a given article---most articles
are read only once, and a click through is only a weak indicator of
whether the article was fully read, let alone liked. Ratings in the
Echo Nest data presumably fall somewhere in between, as the number of
times a user listens to a song likely reveals some indirect
information about their preferences.

As such, we treat each data set as a source of implicit feedback,
where an observed positive rating indicates that a user likes a
particular item, but the rating value itself is ignored. The Mendeley
data are already of this simple binary form. For the Echo Nest and New
York Times data, we consider any song play or article click as a
positive rating, regardless of the play or click count. We also
consider two versions of the Netflix data---the original, explicit
ratings, and an implicit version in which only 4 and 5 star ratings
are retained as observations~\cite{Paquet:2013p9197}.

{\bf Competing methods.} We compare Poisson factorization against an array of
competing methods:
\begin{itemize}
  \item {\bf NMF}: Non-negative Matrix
    Factorization~\cite{Lee:1999}. In NMF, user preferences and item
    attributes are modeled as non-negative vectors in a
    low-dimensional space. These latent vectors are randomly
    initialized and modified via an alternating multiplicative update
    rule to minimize the Kullback-Leibler divergence between the
    actual and modeled rating matrices.

  \item {\bf LDA}: Latent Dirichlet Allocation~\cite{Blei:2003b}. LDA
    is a Bayesian probabilistic generative model where user preferences
    are represented by a distribution over different topics, and each
    topic is a distribution over items. Interest and topic
    distributions are randomly initialized and updated using
    stochastic variational inference~\cite{Hoffman:2013} to
    approximate these intractable posteriors.

  \item {\bf MF}: Probabilistic Matrix Factorization with
    user and item biases. We use a variant of matrix factorization
    popularized through the Netflix Prize~\cite{Koren:2009}, where a
    linear predictor---comprised of a constant term, user activity and
    item popularity biases, and a low-rank interaction term---is fit
    to minimize the mean squared error between the predicted and
    observed rating values, subject to L2 regularization to avoid
    overfitting. Weights are randomly initialized and updated via
    stochastic gradient descent using the Vowpal Wabbit
    package~\cite{Weinberger:2009}. This corresponds to maximum
    a-posteriori inference under Probabilistic Matrix
    Factorization~\cite{Salakhutdinov:2008a}.
\end{itemize}

We note that while HPF, BPF, and LDA take only the non-zero observed
ratings as input, traditional matrix factorization requires that we
provide explicit zeros in the ratings matrix as negative examples for
the implicit feedback setting. In practice, this amounts to either
treating all missing ratings as zeros (as in NMF) and down-weighting
to balance the relative importance of observed and missing
ratings~\cite{Hu:2008p9402}, or generating negatives by randomly
sampling from missing ratings in the training
set~\cite{Gantner:2012p9364,Dror:2012a,Paquet:2013p9197}.  We take the
latter approach for computational convenience, employing a
popularity-based sampling scheme: we sample users by activity---the
number of items rated in the training set---and items by
popularity---the number of training ratings an item received to
generate negative examples.\footnote{We also compared this to a
  uniform random sampling of negative examples, but found that the
  popularity-based sampling performed better.} 

Finally, we note a couple of candidate algorithms that failed to scale
to our data sets. The fully Bayesian treatment of the Probabilistic
Matrix Factorization~\cite{Salakhutdinov:2008}, uses a MCMC algorithm
for inference. The authors~\cite{Salakhutdinov:2008} report that a
single Gibbs iteration on the Netflix data set with 60 latent factors,
requires ~30 minutes, and that they throw away the first 800
samples. This implies at least 16 days of training, while the HPF
variational inference algorithm converges within 13 hours on the
Netflix data. Another alternative, Bayesian Personalized Ranking
(BPR)~\cite{Rendle:2009p9243,Gantner:2012p9364}, optimizes a
ranking-based criteria using stochastic gradient descent. The
algorithm performs an expensive bootstrap sampling step at each
iteration to generate negative examples from the vast set of
unobserved. We found time and space constraints to be prohibitive when
attempting to use BPR with the data sets considered here.


\begin{figure*}[t!]
\centering
\includegraphics[width=\textwidth]{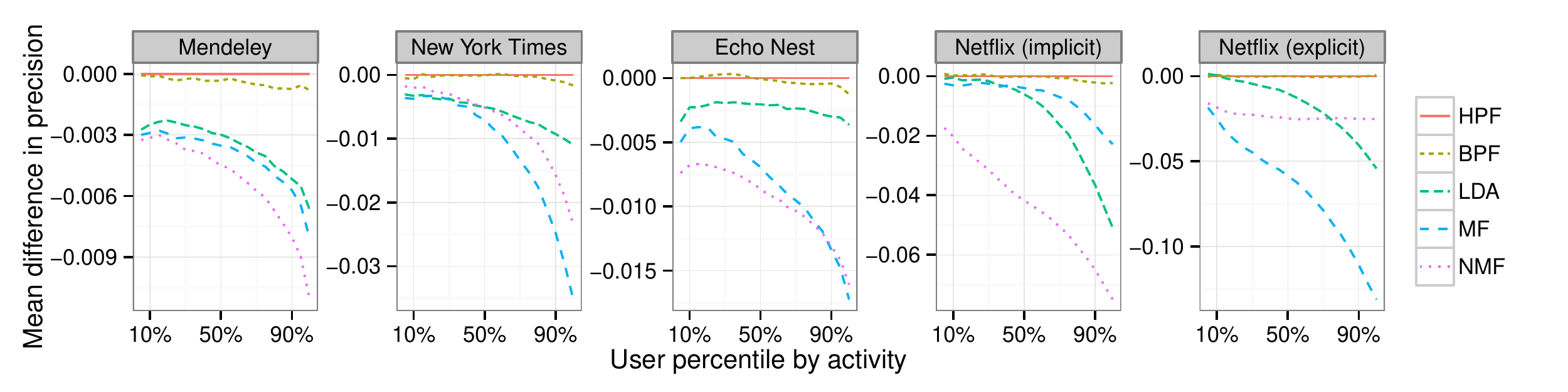}\\
\includegraphics[width=\textwidth]{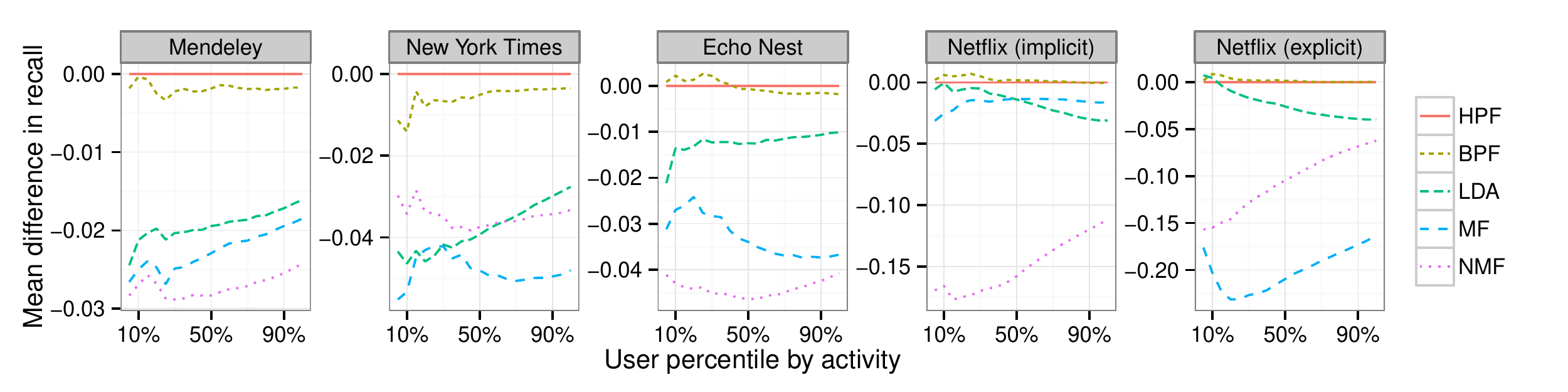}\\
\caption{Predictive performance across users. The top and bottom plots show the
  mean difference in precision and recall to HPF at 20 recommendations,
  respectively, by user activity.}
\label{fig:precision_recall_by_user_activity}
\end{figure*}

{\bf Evaluation.} Prior to training any models, we randomly select
20\% of ratings in each data set to be used as a held-out test set
comprised of items that the user has consumed. Additionally, we set
aside 1\% of the training ratings as a validation set and use it to
determine algorithm convergence and to tune free parameters. We used
the BPF and HPF settings described in \mysec{inference} across all
data sets, and set the number of latent components $K$ to $100$.

During testing, we generate the top $M$ recommendations for each user
as those items with the highest predictive score under each
method. For each user, we compute a variant of precision-at-$M$ that
measures the fraction of relevant items in the user's top-$M$
recommendations. So as not to artificially deflate this measurement
for lightly active users who have consumed fewer than $M$ items, we
compute {\it normalized} precision-at-$M$, which adjusts the
denominator to be at most the number of items the user has in the
test set. Likewise, we compute recall-at-$M$, which captures the
fraction of items in the test set present in the top $M$
recommendations.


\myfig{precision_recall} shows the normalized mean precision at 20
recommendations for each method and data sets. We see that HPF and BPF
outperform other methods on all data sets by a sizeable margin---as
much as 8 percentage points. Poisson factorization provides high-quality
recommendations---a relatively high fraction of items recommended
by HPF are found to be relevant, and many relevant items are
recommended. While not shown in these plots, the relative performance
of methods within a data set is consistent as we vary the number of
recommendations shown to users. We also note that while Poisson
factorization dominates across all of these data sets, the relative
quality of recommendations from competing methods varies substantially
from one data set to the next. For instance, LDA performs quite well on
the Echo Nest data, but fails to beat classical matrix factorization
for the implicit Netflix data set.

We also study precision and recall as a function of user activity to
investigate how performance varies across users of different types. In
particular, \myfig{precision_recall_by_user_activity} shows the mean difference
in precision and recall to HPF, at 20 recommendations, as we look at
performance for users of varying activity, measured by percentile. For example,
the 10\% mark shows mean performance across the bottom 10\% of users, who are
least active; the 90\% mark shows the mean performance for all but the top 10\%
of most active users. Here we see that Poisson factorization outperforms other
methods for users of all activity levels---both the ``light'' users who
constitute the majority, and the relatively few ``heavy'' users who consume
more---for all data sets.

\begin{figure*}
\centering
\includegraphics[width=\textwidth]{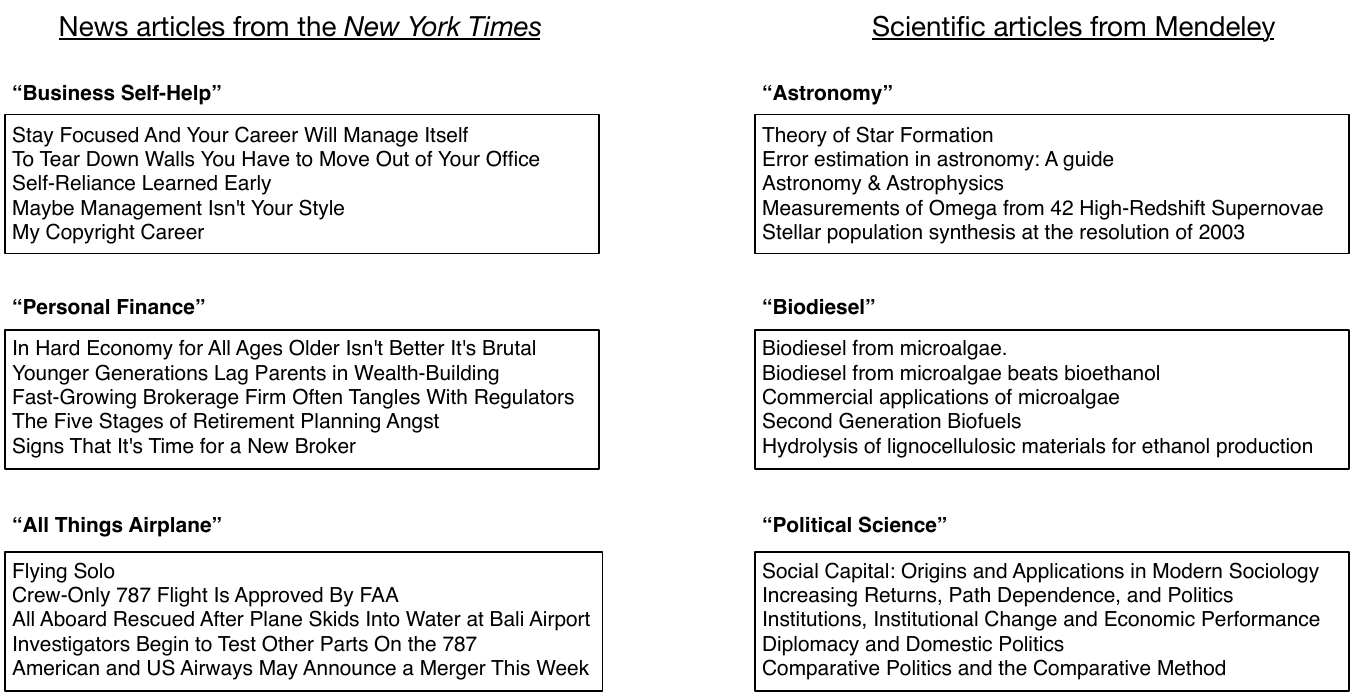}
\caption{The top 10 items by the expected weight $\beta_i$ from three
  of the 100 components discovered by our algorithm for the New York
  Times and Mendeley data sets.}
\label{fig:components}
\end{figure*}

{\bf Exploratory analysis.} The fitted model can be explored to
discover latent structure among items and users and to confirm that
the model is capturing the components in the data in a reasonable
way. For example, in \myfig{components} we illustrate the components
discovered by our algorithm on the scientific articles in Mendeley and
news articles in the New York Times. For each data set, the
illustration shows the top items---items sorted in decreasing order of
their expected weight $\beta_i$---from three of the 100 components
discovered by our algorithm. From these, we see that learned
components both cut across and differentiate between conventional
topics and categories. For instance, in the New York Times data, we find
that multiple business-related topics (e.g., self help and personal
finance) comprise separate components, whereas other articles that
appear across different sections of the newspaper (e.g., business and
regional news) are unified by their content (e.g., airplanes). 


\section{Conclusion}
We have demonstrated that Poisson factorization is an efficient and
effective means of generating high quality recommendations across a
variety of data sets ranging from movie views to scientific article
libraries. It significantly outperforms existing recommendation
methods on both explicit rating data and implicit behavior data,
without the need for ad hoc modifications.  Poisson factorization
algorithms scale to massive data and differ from traditional methods
in their ability to capture the heterogeneity amongst users and items,
accounting for the wide range of activity and popularity amongst them,
respectively.


Future work includes extensions to HPF, to provide cold-start recommendations
using text data~\cite{Wang:2011b}, and to infer the number of latent components
using Bayesian nonparametric assumptions~\cite{Zhou:2012}; and stochastic
variational inference~\cite{Hoffman:2013}, to analyze data sets larger than
those we studied.

\bibliographystyle{abbrv}
\bibliography{bib}

\appendix

Given an observed matrix of user behavior $y$, we would like to
compute the posterior distribution of user preferences $\theta_{uk}$,
item attributes $\beta_{ik}$, user activity $\xi_u$ and item
popularity $\eta_i$, $p(\theta, \beta, \xi, \eta \g y)$.  Our
derivation of the variational algorithm for HPF makes use of
general results about the class of \textit{conditionally conjugate}
models~\cite{Ghahramani:2001,Hoffman:2013}.  We define the class, show
that HPF is in the class, and then derive the variational
inference algorithm.

{\bf Complete conditionals.}  Variational inference fits the
variational parameters to minimize their KL divergence to the
posterior. For the large class of conditionally conjugate models, we
can easily perform this optimization with a coordinate-ascent
algorithm, one in which we iteratively optimize each variational
parameter while holding the others fixed.  A \textit{complete
conditional} is the conditional distribution of a latent variable
given the observations and the other latent variables in the model.  A
conditionally conjugate model is one where each complete conditional
is in an exponential family.

HPF, with the $z_{ui}$ variables described in \mysec{inference}, is a
conditionally conjugate model.  (Without the auxiliary variables, it
is not conditionally conjugate.) For the user weights $\theta_{uk}$,
the complete conditional is a Gamma,
\begin{equation}
  \label{eq:user-weight-cc}
  \theta_{uk} \g \beta, \xi, z, y \sim
  \gam(a + \textstyle \sum_{i} z_{uik}, \xi_u + \sum_{i} \beta_{ik}).
\end{equation}
The complete conditional for item weights $\beta_{ik}$ is symmetric,
\begin{equation}
  \label{eq:item-weight-cc}
  \beta_{ik} \g \theta, \eta, z, y \sim
  \gam(a + \textstyle \sum_{u} z_{uik}, \eta_i + \sum_{i} \theta_{uk}).
\end{equation}
These distributions stem from conjugacy properties between the Gamma
and Poisson. In the user weight distribution, for example, the item
weights $\beta_{ik}$ act as ``exposure'' variables~\cite{Gelman:1995}.
(The roles are reversed in the item weight distribution.) We can
similarly write down the complete conditionals for the user activity
$\xi_u$ and the item popularity $\eta_i$.
\begin{align*}
  \label{eq:user-weight-cc}
  \xi_{u} \g \theta \sim
  \gam(a' + \textstyle Ka, b' + \sum_{k} \theta_{uk}).\nonumber\\
  \eta_{i} \g \beta \sim
  \gam(c' + \textstyle Kc, d' + \sum_{k} \beta_{ik}).\nonumber\\
\end{align*}
The final latent variables are the auxiliary variables.  Recall that
each $z_{ui}$ is a $K$-vector of Poisson counts that sum to the
observation $y_{ui}$. The complete conditional for this vector is
\begin{equation}
  \label{eq:aux-cc}
  z_{ui} \g \beta, \theta, y \sim \mult\left(y_{ui}, \frac{\theta_{u} 
      \beta_{i}}{\textstyle \sum_{k} \theta_{uk} \beta_{ik}}\right).
\end{equation}
Though these variables are Poisson in the model, their complete
conditional is multinomial.  The reason is that the conditional
distribution of a set of Poisson variables, given their sum, is a
multinomial for which the parameter is their normalized set of
rates. (See ~\cite{Johnson:2005, Cemgil:2009}.)

{\bf Deriving the algorithm.}
We now derive variational inference for HPF. First, we set each
factor in the mean-field family (\myeq{q}) to be the same type of
distribution as its complete conditional.  The complete conditionals
for the item weights $\beta_{ik}$ and user weights $\theta_{uk}$ are
Gamma distributions (Equations \ref{eq:user-weight-cc} and
\ref{eq:item-weight-cc}); thus the variational parameters
$\lambda_{ik}$ and $\gamma_{uk}$ are Gamma parameters, each containing
a shape and a rate.  Similarly, the variational user activity
parameters $\kappa_u$ and the variational item popularity parameter
$\tau_i$ are Gamma parameters, each containing a shape and a rate.
The complete conditional of the auxiliary variables $z_{uik}$ is a
multinomial (\myeq{aux-cc}); thus the variational parameter
$\phi_{ui}$ is a multinomial parameter, a point on the $K$-simplex,
and the variational distribution for $z_{ui}$ is $\mult(y_{ui},
\phi_{ui})$.

In coordinate ascent we iteratively optimize each variational
parameter while holding the others fixed.  In conditionally conjugate
models, this amounts to setting each variational parameter equal to
the expected parameter (under $q$) of the complete conditional.
\footnote{It is a little more complex then this. For details, see~\cite{Hoffman:2013}.}  
The parameter to each complete conditional is a function of the other
latent variables and the mean-field family sets all the variables to
be independent.  These facts guarantee that the parameter we are
optimizing will not appear in the expected parameter.

For the user and item weights, we update the variational shape and
rate parameters. The updates are
\begin{eqnarray}
  \gamma_{uk} &=& \langle a + \textstyle \sum_{i} y_{ui} \phi_{uik},
  b + \textstyle \sum_i \lambda_{ik}^{\shape} / \lambda_{ik}^{\rate} \rangle \\
  \lambda_{ik} &=& \langle c + \textstyle \sum_{u} y_{ui} \phi_{uik},
  d + \textstyle \sum_u \gamma_{ik}^{\shape} / \gamma_{ik}^{\rate} \rangle.
\end{eqnarray}
These are expectations of the complete conditionals in
Equations~\ref{eq:user-weight-cc} and \ref{eq:item-weight-cc}.  In the
shape parameter, we use that the expected count of the $k$th item in
the multinomial is $\E_q[z_{uik}] = y_{ui} \phi_{uik}$. In the rate
parameter, we use that the expectation of a Gamma variable is the
shape divided by the rate.

For the variational multinomial the update is
\begin{equation}
  \phi_{ui} \propto \exp\{\Psi(\gamma_{uk}^\shape) - \log
  \gamma_{uk}^{\rate} + \Psi(\lambda_{ik}^\shape) - \log
  \lambda_{ik}^\rate\},
\end{equation}
where $\Psi(\cdot)$ is the digamma function (the first derivative of
the log $\Gamma$ function).  This update comes from the expectation of
the log of a Gamma variable, for example $\E_q[\log \theta_{uk}] =
\Psi(\gamma_{nk}^\shape) - \log \gamma_{nk}^{\rate}$.

\end{document}